\documentclass[a4paper,12pt]{article}
\usepackage{epsfig}
\usepackage[dvips,usenames]{color}
\usepackage{graphicx}

\newlength{\dinwidth}
\newlength{\dinmargin}
\newcommand{\spur}[1]{\not\! #1 \,}
\begin{document}

\title{Charmless $\overline{B}_s\to VV$ decays in QCD factorization}
\bigskip

\author{Xinqiang Li,~ Gongru Lu,~Y.D. Yang\footnote{ Corresponding
author. E-mail address: yangyd@henannu.edu.cn}
\\
{ \small \it Department of Physics, Henan Normal University,
Xinxiang, Henan 453002,  P.R. China}\\
}
\maketitle
\begin{picture}(0,0)
%\put(305,290){\sf hep-ph/0412xxx}
\end{picture}

\bigskip\bigskip
\maketitle

\begin{abstract}
The two body charmless decays of $B_s$ meson to  light vector
mesons are analyzed within the framework of QCD factorization.
This approach implies that the nonfactorizable corrections to
different helicity amplitudes are not the same. The effective
parameters $a_i^h$ for helicity $h=0,+,-$ states receive different
nonfactorizable contributions and hence are helicity dependent,
contrary to naive factorization approach where $a_i^h$ are
universal and polarization independent. The branching ratios for
$\bar{B}_s\to VV$ decays are calculated  and we find that
branching ratios of some channels are of order $10^{-5}$, which
are measurable at future experiments. The transverse to total
decay rate $\Gamma_T/\Gamma$ is also evaluated and found to be
very small for most decay modes, so, in charmless
$\overline{B}_s\to VV$ decays, both light vector mesons tend to
have zero helicity.
\\ {\bf PACS Numbers 13.25.Hw, 12.38.Bx, 12.15Mm}
\end{abstract}

\newpage
\section{ Introduction}

The charmless two-body $B$ decays play a crucial role in
determining the flavor parameters, especially the
Cabibbo-Kobayashi-Maskawa(CKM) angels $\alpha$, $\beta$ and
$\gamma$. With precise measurements of these parameters, we can
explore CP violation which is described by the phase of the CKM
matrix in the standard model(SM). Recently there have been
remarkable progresses in the study of exclusive charmless $B$
decays, both experimentally and theoretically. On the experimental
aspect, many two-body non-leptonic charmless $B$ decays have been
observed by CLEO and $B$-factories at KEK and SLAC
~\cite{barbar,belle} and more $B$ decay channels will be measured
with great precision in the near future. With the accumulation of
data, SM can be tested in more detail. Theoretically, several
novel methods have also been proposed to study the nonfactorizable
effects in the hadronic matrix elements, such as QCD
factorization(QCDF)~\cite{BBNS}, the perturbative QCD
method(pQCD)~\cite{pqcd} and so on. Intensive investigations on
hadronic charmless two-body $B_{u,d}$ decays with these methods
have been studied in detail~\cite{BBNS1,dsd,cheng,MY,pqcd1}.

The extension of QCDF from $B_{u,d}$ decays to $B_s$ decays has
also been carried out by several authors~\cite{hyc,jfs}. In
principle, the physics of the $B_s$ two-body hadronic decays is
very similar to that for the $B_d$ meson, except that the
spectator d quark is replaced by the s quark. However, the problem
is that $B_s$ meson oscillates at a high frequency, and
nonleptonic $B_s$ decays have still remained elusive from
observation. Unlike the $B_{u,d}$ mesons, the heavier $B_s$ meson
cannot be studied at the B-factories operating at the
$\Upsilon(4s)$ resonance. But it is believed that in the
forthcoming hadron colliders such as the Collider Detector at
Fermilab~(CDF), D0, DESY $ep$ collider HERA-B, BTeV, and CERN
Large Hadron Collider~(LHCb), CP violation in $B_s$ system can be
observed with high accuracy. This makes the search for CP
violation in the $B_s$ system decays very interesting.

In the papers~\cite{jfs,BT}, the authors have studied
systematically the $B_s\to PP,PV$ decays(here $P$, $V$ denote
pseudoscalar and vector mesons respectively) with QCD
factorization, and intensive phenomenological analysis has been
made. Since the $B\to VV$ modes reveal dynamics of exclusive $B$
meson decays more than the the $B\to PP$ and $PV$ modes through
the measurement  of the magnitudes and the phases of various
helicity amplitudes, in the present work we plan to make a detail
study of $\overline{B}_s\to VV$ decays within the same framework
of QCD factorization. We find that, contrary to the generalized
factorization approach\cite{hyc}, nonfactorizable corrections to
each helicity amplitude are not the same; the effective parameters
$a_i^h$ vary for different helicity amplitudes. The transverse to
total decay rate $\Gamma_T/\Gamma$ is very small for most decay
modes, so in the heavy quark limit, both light vector mesons in
charmless $\bar{B}_s\to VV$ decays tend to have zero helicity.
Branching ratios for some decay modes are found of order
$10^{-5}$, which could be measured at LHCb.

This paper is organized as follows. In Sec. II, we outline the
necessary ingredients of the QCD factorization approach for
describing the $\bar{B}_s\to VV$  decays and calculate the
effective parameters $a_i^h$. Input parameters, numerical
calculations and results are presented in Sec. III. Finally we
conclude with a summary in Sec. IV. The  amplitudes for charmless
two-body $\overline{B}_s\to VV$ decays are given in Appendix.

\section{$\overline{B}_s\to VV$ in QCD
factorization approach }

\subsection{ The effective Hamiltonian }

Using the operator product expansion and renormalization group
equation, the low energy effective Hamilization relevant to
hadronic charmless B decays can be written as~\cite{coeff}
 \begin{eqnarray}
 {\cal
 H}_{eff}=\frac{G_F}{\sqrt{2}}[\lambda_u(C_1O_1^u+C_2O_2^u)
 +\lambda_c(C_1O_1^c+C_2O_2^c)\nonumber \\
 -\lambda_t(\sum_{i=3}^{10}C_iO_i+C_{7\gamma}O_{7\gamma}+C_{8g}O_{8g})]+h.c
 \end{eqnarray}
where $\lambda_i=V_{ib}V_{iq}^*$ are CKM factors and $C_i({\mu})$
are the effective Wilson coefficients which have been reliably
evaluated to the next-to-leading logarithmic order. The effective
operators $O_i$ can be expressed as follows:
\begin{equation}
\begin{array}{llllll}
O_1^u&=&(\bar{u}b)_{V-A}(\bar{q}u)_{V-A},&
O_2^u&=&(\bar{u}_\alpha b_\beta)_{V-A}(\bar{q}_\beta u_\alpha)_{V-A},\\
O_1^c&=&(\bar{c}b)_{V-A}(\bar{q}c)_{V-A},&
O_2^c&=&(\bar{c}_\alpha b_\beta)_{V-A}(\bar{q}_\beta c_\alpha)_{V-A},\\
O_{3(5)}&=&(\bar{q}b)_{V-A}\sum\limits_{q'}(\bar{q'}q')_{V-A(V+A)},&
O_{4(6)}&=&(\bar{q}_\alpha b_\beta)
_{V-A}\sum\limits_{q'}(\bar{q'}_\beta q'_\alpha)_{V-A(V+A)},\\
O_{7(9)}&=&\frac{3}{2}(\bar{q}b)_{V-A}\sum\limits_{q'}e_{q'}(\bar{q'}q')
_{V+A(V-A)},& O_{8(10)}&=&\frac{3}{2}(\bar{q}_\alpha b_\beta)
_{V-A}\sum\limits_{q'}e_{q'}(\bar{q'}_\beta q'_\alpha)_{V+A(V-A)},\\
O_{7\gamma}&=&\frac{e}{8\pi^2}m_b\bar{q}_\alpha\sigma^{\mu\nu}
 (1+\gamma_5)b_{\alpha}F_{\mu\nu},&
O_{8g}&=&\frac{g}{8\pi^2}m_b\bar{q}_\alpha\sigma^{\mu\nu}
 (1+\gamma_5)T_{\alpha\beta}^{a}b_{\beta}G_{\mu\nu}^{a}.
\end{array}
\end{equation}
Where  $q=d,s$ and $q'$ denotes all the active quarks at the scale
$\mu={\cal O }(m_b)$, i.e., $q'=u,d,s,c,b$.

\subsection{The factorizable amplitude for $\overline{B}_s\to VV$ }

To calculate the decay rate and branching ratios for
$\overline{B}_s\to VV$ decays, we need the hadronic matrix element
for the local four fermion operators
 \begin{equation}
  \langle V_1(\lambda_1)V_2(\lambda_2)|(\bar{q}_2q_3)_{V-A}
(\bar{q}_1b)_{V-A}|\overline{B}_s \rangle,
 \end{equation}
where $\lambda_1$, $\lambda_2$ are the helicities of the
final-state vector mesons $V_1$ and $V_2$ with four-momentum $p_1$
and $p_2$, respectively. In the rest frame of $B_s$ system, since
$B_s$ meson has spin zero, we have $\lambda_1=\lambda_2=\lambda$.
Let $X^{(B_sV_1,V_2)}$ denote the factorizable amplitude with the
vector meson $V_2$ being factored out, under the naive
factorization(NF) approach, we can express $X^{(B_sV_1,V_2)}$ as
\begin{equation}
 X^{(B_sV_1,V_2)}=\langle V_2|(\bar{q}_2q_3)_{V-A}|0\rangle
 \langle V_1|(\bar{q}_1b)_{V-A}|\overline{B}_s \rangle.
\end{equation}
In term of the decay constant and form factors defined
by~\cite{form,MT,Gerard}
\begin{eqnarray}
\langle V(p,\varepsilon^{\ast})|\bar{q}\gamma_{\mu}q'|0
\rangle&=&-i
f_{V}m_V\varepsilon_{\mu}^{\ast},\\
<V(p,\varepsilon^{\ast})|\bar{q}\gamma_{\mu}(1-\gamma_5)b|\overline{B}_s
(p_{B})> &=&-\varepsilon_{\mu}^\ast(m_B+m_V)A_1^{B_sV}(q^2)
+(p_B+p)_{\mu}({\varepsilon^\ast}\cdot{p_B})\frac{A_2^{B_sV}(q^2)}{m_B+m_V}\nonumber \\
&&+q_{\mu}({\varepsilon^\ast}\cdot{p_B})\frac{2m_V}{q^2}[A_3^{B_sV}(q^2)-A_0^{B_sV}(q^2)]\nonumber\\
&&-i\epsilon_{\mu\nu\alpha\beta}\varepsilon^{\ast\nu}p_B^{\alpha}p^{\beta}
\frac{2V^{B_sV}(q^2)}{m_B+m_V},
\end{eqnarray}
where $q=p_B-p$ and the form factors obey the following exact
relations
\begin{eqnarray}
A_3(0)&=&A_0(0)\nonumber,\\
A_3^{B_sV}(q^2)&=&\frac{m_B+m_V}{2m_V}A_1^{B_sV}(q^2)-\frac{m_B-m_V}{2m_V}A_2^{B_sV}(q^2).
\end{eqnarray}
With above equations, the factorizable amplitude for
$\overline{B}_s\to V_1V_2$ can be written as
\begin{eqnarray}
X^{(\overline{B}_s V_1,V_2)}&=&i f_{V_2}m_{V_2} \left[
(\varepsilon_1^{\ast}\cdot\varepsilon_2^{\ast})
(m_{B_s}+m_{V_1})A_1^{B_sV_1}(m_{V_2}^2)-(\varepsilon_1^{\ast}\cdot
p_B)(\varepsilon_2^{\ast}\cdot p_B)
\frac{2A_2^{B_sV_1}(m_{V_2}^2)}{m_{B_s}+m_{V_1}}
\right.\nonumber\\
&&\left.+i\epsilon_{\mu\nu\alpha\beta}\varepsilon_2^{\ast\mu}
\varepsilon_1^{\ast\nu}p_B^{\alpha}p^{\beta}
\frac{2V^{B_sV}(q^2)}{m_{B_s}+m_{V_1}}\right] ,\label{xbs}
\end{eqnarray}
where $p_B(m_{B_s})$ is the four-momentum(mass) of the
$\overline{B}_s$ meson, $m_{V_1}(\varepsilon_1^{\ast})$ and
$m_{V_2}(\varepsilon_2^{\ast})$ are the masses(polarization
vectors) of the two vector mesons $V_1$ and $V_2$ respectively.
Here and in the following throughout the paper we use the sign
convention $\epsilon^{0123}=-1$. Assuming the $V_1$($V_2$) meson
flying in the plus(minus) z-direction carrying the momentum
$p_1$($p_2$), we get
 \begin{equation}
 X^{(\overline{B}_s V_1,V_2)}= \left \{\begin{array}{ll}\frac{-i
f_{V_2}}{2m_{V_1}}\left[(m^2_{B_s}-m_{V_1}^2-m_{V_2}^2)(m_{B_s}+m_{V_1})
A_1^{B_sV_1}(m_{V_2}^2)\right.\nonumber\\ \left.-\frac{4m_{B_s}^2
p_c^2}{m_{B_s}+m_{V_1}}A_2 ^{B_sV_1}(m_{V_2}^2)\right]\equiv h_0 &
{\rm for\ } \lambda=0, \\-i
 f_{V_2}m_{V_2}[(m_{B_s}+m_{V_1})A_1 ^{B_sV_1}(m_{V_2}^2)\pm
\frac{2m_{B_s} p_c}{m_{B_s}+m_{V_1}}V ^{B_sV_1}(m_{V_2}^2)]\equiv
h_{\pm}& {\rm for\ } \lambda={\pm},
 \end{array}
 \right. \label{helicity}
 \end{equation}
where $\lambda=0,{\pm}$ is the helicity of the vector meson and
$p_c=|\vec{p}_1|=|\vec{p}_2|$ is the momentum of either of the two
outgoing vector mesons  in the $\overline{B}_s$ rest frame.

In general, the $\overline{B}_s\to V_1 V_2$ amplitude can be
decomposed into three independent helicity amplitudes $H_0$, $H_+$
and $H_-$, corresponding to $\lambda=0$, $+$ and $-$ respectively.
We use the notation
 \begin{equation}
 H_{\lambda}=<V_1(\lambda)V_2(\lambda)|{\cal H}_{eff}|
\overline{B}_s>
 \end{equation}
for the helicity matrix element and it can be expressed by three
independent $Lorentz$ scalars $a$, $b$ and $c$. The relations
between them can be written as~\cite{BBNS,PV}
 \begin{equation}
 H_{\lambda}=\varepsilon_{1 \mu}^{\ast}\varepsilon_{2
\nu}^{\ast}\left(a
g^{\mu\nu}+\frac{b}{m_{V_1}m_{V_2}}p_B^{\mu}p_B^{\nu}+\frac{i
c}{m_{V_1}m_{V_2}}\epsilon^{\mu\nu\alpha\beta}p_{1 \alpha}p_{2
\beta}\right),
 \end{equation}
where the coefficient $c$ corresponds to the p-wave amplitude, $a$
and $b$ to the mixture of s- and d-wave amplitudes. The helicity
amplitudes can be reconstructed as
\begin{eqnarray}
H_0&=&-a x-b (x^2-1),\\
H_{\pm}&=&-a \mp c \sqrt {x^2-1},
\end{eqnarray}
where $x=\frac{(p_1\cdot p_2)}{(m_{V_1}m_{V_2})}$. Given the
helicity amplitudes, the decay rate and the branching ratio for
$\overline{B}_s\to V_1 V_2$ can be written as
 \begin{eqnarray}
 \Gamma(\overline{B}_s\to V_1 V_2)&=&\frac{p_c}{8\pi
 m_{B_s}^2}\left(|H_0|^2+|H_+|^2+|H_-|^2\right)s,\nonumber\\
 Br(\overline{B}_s\to V_1 V_2)&=&\tau_{B_s}\frac{p_c}{8\pi
 m_{B_s}^2}\left(|H_0|^2+|H_+|^2+|H_-|^2\right)s, \label{ga br}
 \end{eqnarray}
with $s=1/2$ for two identical final states and $s=1$ for the
other cases. where $\tau_{B_s}$ is the lifetime of the $B_s$
meson, and $p_c$ is given by
\begin{equation}
p_c=\frac{1}{2m_{B_s}}\sqrt{[m_{B_s}^2-(m_{V_1}
+m_{V_2})^2][m_{B_s}^2-(m_{V_1}-m_{V_2})^2]}~.
\end{equation}

\subsection{ QCD factorization for $\bar{B}_s\to VV$}

Under the naive factorization(NF) approach, the coefficients $a_i$
are given by $a_i=C_{i}+\frac{1}{N_C}C_{i+1}$ for odd i and $
a_{i}=C_{i}+\frac{1}{N_C}C_{i-1}$ for even i,  which are obviously
independent of the helicity $\lambda$. In the present paper, we
will compute the nonfactorizable corrections to the effective
parameters $a_i^h$, which however are not the same for different
helicity amplitudes $H_0$ and $H_{\pm}$.

The QCD-improved factorization(QCDF) approach advocated by Beneke
et al.~\cite{BBNS} allows us to compute the nonfactorizable
corrections to the hadronic matrix elements $\langle V_1
V_2|O_i|\overline{B}_s \rangle$ in the heavy quark limit, since in
the $m_b\to \infty$ limit only hard interactions between the
($\bar{B}_s V_1$) system and $V_2$ survive. In this method, the
light-cone distribution amplitudes(LCDAs) play an essential role.
Since we are only concerned with two light vector mesons in the
final states, the LCDAs of the light vector meson of interest in
momentum configuration are given by~\cite{MT,PV}
\begin{equation}
{\cal M}_{\delta\alpha}^V={\cal M}_{\delta\alpha\parallel}^V+{\cal
M}_{\delta\alpha\perp}^V
\end{equation}
with (here we suppose the vector meson moving in the
$n_-$-direction)
\begin{eqnarray}
{\cal M}^V_{\parallel}&=&- \frac{i f_V}{4}\frac{m_V
(\varepsilon^{\ast} \cdot n_+)}{2}\spur{n_-} \Phi^V_{\parallel}
(u),\nonumber\\
{\cal M}^V_{\perp}&=&-\frac{i
f_V^{\perp}}{4}E\spur{\varepsilon_{\perp}^{\ast}}
\spur{n_-}\Phi^V_{\perp}(u)-\frac{i f_V
m_V}{4}\left[\spur{\varepsilon_{\perp}^{\ast}} g_{\perp}^{(v)V
}(u)+i \epsilon_{\mu\nu\rho\sigma}
{\varepsilon_{\perp}^{\ast}}^{\nu} n_-^{\rho} n_+^{\sigma}
\gamma^{\mu} \gamma_5 \frac{g_{\perp}^{\prime(a)V}
(u)}{8}\right],\nonumber\\ \label{paral}
\end{eqnarray}
where $n_{\pm}=(1,0,0,\pm 1)$ are the light-cone null vectors, $u$
is the light-cone momentum fraction of the quark in the vector
meson, $f_V$ and $f_V^{\perp}$ are vector and tensor decay
constants, and $E$ is the energy of the vector meson in the $B_s$
rest system. In Eq.(\ref{paral}), $\Phi^V_{\parallel} (u)$ and
$\Phi^V_{\perp} (u)$ are leading-twist distribution
amplitudes(DAs), while $g^{(v)V}_{\perp} (u)$ and
$g_{\perp}^{\prime(a)V}(u)=\frac{{\rm d}g_{\perp}^{(a)V}(u)}{{\rm
d} u}$ are twist-3 ones. Since the twist-2 DA $\Phi_{\perp}^V (u)$
contributions to the vertex corrections and hard spectator
interactions vanish in the chiral limit, and furthermore, the
contributions of the twist-3 DAs $h_{\parallel}^{(s,t)}(u)$ are
power suppressed compared to that of the leading twist ones for
the helicity zero case, therefore we will work to the
leading-twist approximation for longitudinally polarized states
and to the twist-3 level for transversely polarized ones. We note
that the same observation has been made by Cheng and
Yang~\cite{chengyang} in studying $B_{u,d}\to \phi K^*$.

In the heavy quark limit, the light-cone projector for $B$ meson
can be expressed as~\cite{BBNS1,grozin}
 \begin{equation}
 {\cal M}_{\alpha\beta}^B=-\frac{i f_B m_B}{4}\left[(1+\spur{v}
)\gamma_5 \left\{\Phi_1^B (\xi) +\spur{n_-} \Phi_2^B
(\xi)\right\}\right]_{\beta\alpha},
 \label{projector}
 \end{equation}
with the normalization condition
 \begin{equation}
 \int_{0}^{1}{\rm d}{\xi}\Phi_1^B (\xi)=1,~~~~~\int_{0}^{1}{\rm
d}{\xi}\Phi_2^B (\xi)=0,
 \end{equation}
where $\xi$ is the momentum fraction of the spectator quark in the
$B$ meson.

Equipped with these preliminaries, we can now calculate the
nonfactorizable  corrections to the effective parameters $a_i^h$
systematically. After  direct calculations, we get
\begin{eqnarray}
a_1^h &=&C_1+\frac{C_2}{N_C}+\frac{\alpha_s}{4\pi}\frac{C_F}{N_C}C_2(f_I^h+f_{II}^h),\nonumber\\
a_2^h &=&C_2+\frac{C_1}{N_C}+\frac{\alpha_s}{4\pi}\frac{C_F}{N_C}C_1(f_I^h+f_{II}^h),\nonumber\\
a_3^h &=&C_3+\frac{C_4}{N_C}+\frac{\alpha_s}{4\pi}\frac{C_F}{N_C}C_4(f_I^h+f_{II}^h),\nonumber\\
a_4^h &=&C_4+\frac{C_3}{N_C}+\frac{\alpha_s}{4\pi}\frac{C_F}{N_C}C_3(f_I^h+f_{II}^h)\nonumber\\
&&+\frac{\alpha_s}{4\pi}\frac{C_F}{N_C}
\left\{(C_3-\frac{1}{2}C_9)[G^h
(s_q)+G^h(s_b)-\left(\begin{array}{c}\frac{4}{3}\\
\frac{2}{3}\end{array}\right)]\right.\nonumber\\
& &-C_1 [\frac{\lambda_u}{\lambda_t}G^h (s_u)+\frac{\lambda_c}{\lambda_t} G^h (s_c)+\left(\begin{array}{c}\frac{2}{3}\\
\frac{1}{3}\end{array}\right)]\nonumber\\
& &+\left.(C_4+C_6)\sum_{i=u}^b G^h
(s_i)+\frac{3}{2}(C_8+C_{10})\sum_{i=u}^b G^h(s_i)+C_{8g} G_g^h
\right\}, \nonumber\\
a_5^h &=&C_5+\frac{C_6}{N_C}-\frac{\alpha_s}{4\pi}\frac{C_F}{N_C}C_6(f_I^h+f_{II}^h),\nonumber\\
a_6^h &=&C_6+\frac{C_5}{N_C},\nonumber\\
a_7^h &=&C_7+\frac{C_8}{N_C}-\frac{\alpha_s}{4\pi}\frac{C_F}{N_C}C_8(f_I^h+f_{II}^h),\nonumber\\
a_8^h &=&C_8+\frac{C_7}{N_C},\nonumber\\
a_9^h &=&C_9+\frac{C_{10}}{N_C}+\frac{\alpha_s}{4\pi}\frac{C_F}{N_C}C_{10}(f_I^h+f_{II}^h),\nonumber\\
a_{10}^h&=&C_{10}+\frac{C_9}{N_C}+\frac{\alpha_s}{4\pi}\frac{C_F}{N_C}C_9(f_I^h+f_{II}^h),
\label{coeff}
\end{eqnarray}
where $C_F=(N_C^2-1)/{2 N_C}$, and $N_C=3$ is the number of
colors, $s_i=m_i^2/m_b^2$ and $q=d, s$(determined  by the $b\to d$
or $b\to s$ transition process). The superscript $h$ denotes the
polarization of the vector meson(which is equivalent to $\lambda$,
but for convenience we shall adopt $h$ in the following) where
$h=0$ denotes helicity $0$ state and $h={\pm}$ for helicity $\pm$
ones. In the expression $a_4^h$, the upper value in parenthesis
corresponds to $h=0$ state, while the lower value to $h={\pm}$
ones.

In Eq.(\ref{coeff}), $f_I^h$ denotes the contributions from the
vertex corrections. In the naive dimensional regularization(NDR)
scheme for $\gamma_5$, it is given by
 \begin{eqnarray}
 f_I^0&=&-12\log \frac{\mu}{m_b}-18+\int_0^1{\rm d}u
\Phi_{\parallel}^{V_2} (u)(3\frac{1-2u}{1-u}\log u-3i \pi),\nonumber\\
 f_I^{\pm}&=&-12\log \frac{\mu}{m_b}-16+\int_0^1{\rm d}u
[g_{\perp}^{(v){V_2}} (u)\mp \frac{g_{\perp}^{\prime(a){V_2}}
(u)}{4}\zeta]\left \{3\frac{1-2u}{1-u}\log u-3i \pi)\right.\nonumber\\
&&\left.+2\int_0^1{\rm d}x{\rm d}y[\frac{1-x-y}{x y}-\frac{u}{x
u+y}\mp\frac{(1-x)u}{y(xu+y)}]\right \},
 \end{eqnarray}
where $\zeta=+1$ or $-1$, corresponding to $(V-A)\otimes(V-A)$ or
$(V-A)\otimes(V+A)$ current respectively. It is obvious that
$f_I^0$ has the same expression as the hard scattering kernel
$F_{M_2}$ for $B\to \pi\pi$ mode~\cite{BBNS,MY} as it should be.

For hard spectator interactions, supposing ${V_1}$ be the recoiled
meson and $V_2$ the emitted meson, explicit calculations for
$f_{II}^h$ yields
\begin{eqnarray}
f_{II}^0&=&-\frac{4
\pi^2}{N_C}\frac{if_{B_s}f_{V_1}f_{V_2}}{h_0}\int_0^1{\rm d}\xi
\frac{\Phi_1^B (\xi)}{\xi}\int_0^1{\rm d}v
\frac{\Phi_{\parallel}^{V_1} (v)}{\bar{v}}\int_0^1{\rm d}u
\frac{\Phi_{\parallel}^{V_2}
(u)}{u},\nonumber\\
f_{II}^{\pm}&=&\frac{4
\pi^2}{N_C}\frac{if_{B_s}f_{V_1}^{\perp}f_{V_2}m_{V_2}}{m_{B_s}h_{\pm}}2(1\pm1)\int_0^1{\rm
d}\xi \frac{\Phi_1^B (\xi)}{\xi}\int_0^1{\rm d}v
\frac{\Phi_{\perp}^{V_1} (v)}{\bar{v}^2}\int_0^1{\rm d}u
(g_{\perp}^{(v){V_2}} (u)\mp \frac{g_{\perp}^{\prime(a){V_2}}
(u)}{4}\zeta)\nonumber\\ &&-\frac{4
\pi^2}{N_C}\frac{if_{B_s}f_{V_1}f_{V_2}m_{V_1}m_{V_2}}{m_{B_s}^2h_{\pm}}\int_0^1{\rm
d}\xi \frac{\Phi_1^B (\xi)}{\xi}\int_0^1{\rm d}v {\rm
d}u(g_{\perp}^{(v){V_1}} (v)\mp \frac{g_{\perp}^{\prime(a){V_1}}
(v)}{4})\nonumber\\&&(g_{\perp}^{(v){V_2}} (u)\mp
\frac{g_{\perp}^{\prime(a){V_2}} (u)}{4}\zeta)\frac{u+\bar v}
{u{\bar v}^2}, \label{hard}
\end{eqnarray}
with $\bar{v}=1-v$, and $h_0, h_{\pm}$ given by
Eq.~(\ref{helicity}). In Eq.~(\ref{hard}), when we adopt the
asymptotical form for the vector meson LCDAs, there will be a
logarithmic infrared divergence with regard to the $v$ integral in
$f_{II}^{\pm}$, which implies that the spectator interaction is
dominated by soft gluon exchanges in the final states. In analogy
with the treatment in works~\cite{BBNS1,dsd,jfs}, we parameterize
it as
 \begin{equation}
 X_h=\int_0^1{\rm
d}x \frac{1}{x} =\log \frac{m_b}{\Lambda_h}(1+\rho_H e^{i\phi_H}),
 \end{equation}
with $(\rho_H,\phi_H)$ related to the contributions from hard
spectator scattering. Since the parameters $(\rho_H,\phi_H)$ are
unknown, how to treat them is a major theoretical uncertainty in
the QCD factorization approach. In the later numerical analysis,
we shall take $\Lambda_h=0.5GeV$,
$(\rho_h,\phi_h)=(0,0)$~\cite{jfs} as our default values.

In calculating the contributions of the QCD penguin-type diagrams,
we should pay attention to the fact that there are two distinctly
different contractions argued in~\cite{BBNS1}. With this in mind,
the nonfactorizable corrections induced by local four-quark
operators $O_i$ can be described by the function $G^h (s)$ which
is given by
\begin{eqnarray}
G^0 (s)&=&-\frac{2}{3}+\frac{4}{3}\log
\frac{\mu}{m_b}-4\int_0^1{\rm d}u ~\Phi_{\parallel}^{V_2}
(u) g(u,s),\nonumber\\
G^{\pm}(s)&=&-\frac{2}{3}+\frac{2}{3}\log
\frac{\mu}{m_b}-2\int_0^1{\rm d}u ~(g_{\perp}^{(v){V_2}}(u)\mp
\frac{g_{\perp}^{\prime(a){V_2}}(u)}{4}) g(u,s),
\end{eqnarray}
with the function
 \begin{equation}
 g(u,s)=\int_0^1{\rm d}x~ x\bar{x}\log{[s-x\bar{x}(1-u)-i\epsilon]}.
 \end{equation}

In Eq.(\ref{coeff}), we also take into account the contributions
of the dipole operator $O_{8g}$ which will give a tree-level
contribution described by the function $G_g^h$ defined as
\begin{eqnarray}
G_g^0&=&\int_0^1 {\rm d}u ~\frac{2\Phi_{\parallel}^{V_2}
(u)}{1-u},\nonumber\\
G_g^+&=&\int_0^1{\rm d}u ~(g_{\perp}^{(v){V_2}}(u)-
\frac{g_{\perp}^{\prime(a){V_2}}(u)}{4}),\nonumber\\
G_g^-&=&\int_0^1{\rm d}u ~(g_{\perp}^{(v){V_2}}(u)-
\frac{g_{\perp}^{\prime(a){V_2}}(u)}{4})\frac{1}{1-u}.
\end{eqnarray}

Due to   $<V|\bar{q}_1q_2 |0>=0$, $\overline{B}_s\to V_1 V_2$
decays do not receive nonfactorizable contributions from $a_6^h$
and $a_8^h$ penguin terms as shown in Eq.(\ref{coeff}).

\section{ Numerical results and discussions }

To proceed, we use the next-to-leading order Wilson coefficients
in the NDR scheme for $\gamma_5~\cite{jfs}$
\begin{eqnarray}
C_1&=&1.078,\ C_2=-0.176,\ C_3=0.014,\ C_4=-0.034,\ C_5=0.008,\ C_6=-0.039,\nonumber\\
C_7/\alpha&=&-0.011,\ C_8/{\alpha}=0.055,\ C_9/{\alpha}=-1.341,\
C_{10}/{\alpha}=0.264,\ C_{8g}=-0.146.
\end{eqnarray}
at $\mu=m_b=4.66$ GeV, with $\alpha$ being the electromagnetic
fine-structure coupling constant. For quark masses, which appears
in the penguin loop corrections with regard to the functions $G^h
(s)$, we take
\begin{equation}
m_u=m_d=m_s=0,\ m_c=1.47~{\rm GeV},\ m_b=4.66~{\rm GeV}.
\end{equation}

As for the CKM matrix elements, we adopt the Wolfenstein
parametrization up to ${\cal O} (\lambda^3)$:
 \begin{equation} \left(\begin{array}{lll}
 V_{ud} & V_{us} & V_{ub}\\
 V_{cd} & V_{cs} & V_{cb}\\
 V_{td} & V_{ts} & V_{tb}\\
 \end{array}\right)
 =\left(\begin{array}{ccc}
 1-\lambda ^2 /2 & \lambda & A\lambda^3(\rho-i \eta)\\
  -\lambda & 1-\lambda^2 /2 & A\lambda^2 \\
 A\lambda^3(1-\rho-i \eta) & -A\lambda^2  & 1\\
 \end{array}\right).
 \end{equation}
for the Wolfenstein parameters appearing in the above expression,
we shall use the values given by~\cite{PDG}
\begin{equation}
\lambda=0.2236,\ A=0.824,\ \bar{\rho}=0.22,\ \bar{\eta}=0.35,
\end{equation}
where $\bar{\rho}=\rho(1-\frac{\lambda^2}{2})$ and
$\bar{\eta}=\eta(1-\frac{\lambda^2}{2})$. For computing the
branching ratio, the lifetime of $B_s$ meson is $\tau_{B_s}=1.461$
ps~\cite{PDG}.

For the LCDAs of the vector meson, we use the asymptotic
form~\cite{PV}
\begin{eqnarray}
\Phi_{\parallel}^V (x)&=&\Phi_{\perp}^V
(x)=g_{\perp}^{(a)V}=6x(1-x),\nonumber\\
g_{\perp}^{(v)V} (x)&=&\frac{3}{4}[1+(2x-1)^2].
\end{eqnarray}
As for the two $B_s$ meson wave functions given by
Eq.(\ref{projector}), we find that only $\Phi^B_1 (\xi)$ has
contributions to the nonfactorizable corrections. We adopt the
moments of the $\Phi_1^B (\xi)$ defined by ~\cite{BBNS,BBNS1} in
our numerical evaluation
\begin{equation}
\int_0^1{\rm d}\xi\frac{\Phi_1^B
(\xi)}{\xi}=\frac{m_{B_s}}{\Lambda_B},
\end{equation}
with $\Lambda_B=0.35$~GeV. The quantity $\Lambda_B$ parameterizes
our ignorance about the $B_s$ meson distribution amplitudes and
thus  brings large theoretical uncertainty.

The decay constants and form factors are nonperturbative
parameters which are taken as input parameters. In principle, they
are available from the experimental data and /or estimated with
well-founded theories, such as lattice calculations, QCD sum rules
etc. For the decay constants, we take their values in our
calculations as~\cite{BBNS1,jfs, cdlu, form}
\begin{eqnarray}
&&f_{B_s}=236~{\rm MeV},\ f_{K^{\ast}}=214~{\rm MeV}, \
f_{K^{\ast}}^{\perp}=175~{\rm MeV}, \ f_{\rho}=210~{\rm MeV},
\nonumber
\\ \ &&f_{\omega}=195~{\rm MeV},\ f_{\phi}=233~{\rm MeV}, \
f_{\phi}^{\perp}=175~{\rm MeV}.
\end{eqnarray}
For the form factors involving the $B_s\to K^{\ast}$ and
$B_s\to\phi$ transition, we adopt the results given by~\cite{form}
which are analyzed using the light-cone sum rule(LCSR) method with
the parameterization
\begin{equation}
f(q^2)=\frac{f(0)}{1-a_F(q^2/m_{B_s}^2)+b_F(q^2/m_{B_s}^2)^2}
\end{equation}
for the form-factor $q^2$ dependence. At the maximum recoil, the
form factors are listed as~\cite{form}
 \begin{eqnarray}
 A_1^{B_s\phi}(0)&=&0.296,~~~a_F=0.87,~~~b_F=-0.061,\nonumber\\
 A_2^{B_s\phi}(0)&=&0.255,~~~a_F=1.55,~~~b_F=0.513,\nonumber\\
 V^{B_s\phi} (0) &=&0.433,~~~a_F=1.75,~~~b_F=0.736,\nonumber\\
 A_1^{B_sK^{\ast}}(0)&=&0.190,~~~a_F=1.02,~~~b_F=-0.037,\nonumber \\
 A_2^{B_sK^{\ast}}(0)&=&0.164,~~~a_F=1.77,~~~b_F=0.729,\nonumber\\
 V^{B_sK^{\ast}} (0) &=&0.262,~~~a_F=1.89,~~~b_F=0.846.
 \end{eqnarray}
It is obvious that the $q^2$ dependence for the form factors $A_2$
and $V$ are dominated by the dipole terms, while $A_1$ by the
monopole term in the region where $q^2$ is not too large.

To illustrate the non-universality of the nonfactorizable effects
on different helicity amplitudes, we list a few numerical results
of the parameters $a_i^h$ for a specific mode $\overline{B}_s\to
K^{+\ast}\rho^-$ in Table 1.  In order to compare with the
parameters $a_i$ in the NF approach, we also present the results
of $a_i$ calculated in NF approach.

%the section of table:
\begin{table}[htbp]
\caption{
 The effective parameters $a_i^h$ in the NF and QCDF approach for $\bar{B}_s\to
K^{+\ast}\rho^-$. \vspace{0.1in} }
\begin{center}
\doublerulesep 0.8pt \tabcolsep 0.1in
\begin{tabular}{ccc}\hline\hline
$a_i^h$ &NF& QCDF\\
\hline $a_1^0$& $1.0193$&$1.0265+0.0126 i$\\
$a_4^0$& $-0.0293$&$-0.0263-0.0015 i$\\
 $a_{10}^0 $& $-0.0013$&$-0.0009+0.0007 i$\\
\hline $a_1^+$& $1.0193$&$1.0701+0.0126 i$\\
 $a_4^+$& $-0.0293$&$-0.0385-0.0015 i$\\
 $a_{10}^+ $& $-0.0013$&$0.0015+0.0007 i$\\
\hline $a_1^-$& $1.0193$&$1.0943+0.0126 i$\\
 $a_4^-$& $-0.0293$&$-0.0374+0.0022 i$\\
 $a_{10}^-$& $-0.0013$&$0.0028+0.0007 i$\\
\hline\hline
\end{tabular}
\end{center}
\end{table}
From Table 1, we can see that nonfactorizable corrections to the
helicity amplitudes are not universal. The effective parameters
$a_i^h$ for helicity $h=0,+,-$ states receive different
nonfactorizable contributions and hence they are helicity
dependent, quite contrary to the naive factorization(NF) approach
where the parameters $a_i$ are universal and polarization
independent.

The branching ratios for several channels of $\overline{B}_s\to
VV$ decays in the LCSR analysis for form factors are collected in
Table 2. In order to compare the size of different helicity
amplitudes, we define two quantities:
\begin{eqnarray}
\frac{\Gamma_T}{\Gamma}&=&\frac{|H_+|^2+|H_-|^2}{|H_0|^2+|H_+|^2+|H_-|^2},\\
\frac{\Gamma_L}{\Gamma}&=&\frac{|H_+|^2+|H_-|^2}{|H_0|^2+|H_+|^2+|H_-|^2}.
\end{eqnarray}
The ratios of $\Gamma_T/ \Gamma$ and $\Gamma_L/ \Gamma$ measure
the relative amount of transversely and longitudinally polarized
vector meson. In Table II, we also give the values of
$\Gamma_T/\Gamma$ for each channel both in  QCD
factorization(QCDF) approach and the naive factorization(NF)
approach.

%the section of table:
\begin{table}[htbp]
\caption{  Branching ratios and the transverse to total decay rate
$\Gamma_T/\Gamma$ for charmless $\bar{B}_s\to VV$ decays in QCD
factorization(QCDF) approach and in the NF approach.}
\begin{center}
\doublerulesep 0.8pt \tabcolsep 0.1in
\begin{tabular}{lcccc}\hline\hline
 \multicolumn{2}{c@{\hspace{-6cm}}}{$\Gamma_T/\Gamma$} & \multicolumn{2}{c@{\hspace{-4cm}}}{BR } \\
\cline{2-3}\cline{4-5}\raisebox{2.3ex}[0pt]{channel}& QCDF & NF
& QCDF  & NF \\
\hline
$b\to d$ transition\\
 $\overline{B}_s\to K^{+\ast}\rho^-$& $0.071$&$0.066$&$1.82\times
 10^{-5}$&$1.79\times
 10^{-5}$\\
 $\overline{B}_s\to K^{0\ast}\rho^0$& $0.072$&$0.062$&$5.29\times
 10^{-7}$&$5.94\times10^{-7}$\\
 $\overline{B}_s\to K^{0\ast}\omega$& $0.046$&$0.064$&$3.08\times
 10^{-7}$&$7.32\times 10^{-7}$\\\hline
 $b\to s$ transition\\
 $\overline{B}_s\to K^{+\ast}K^{-\ast}$&
 $0.100$&$0.103$&$1.94\times 10^{-6}$&$1.58\times 10^{-6}$\\
 $\overline{B}_s\to\omega\phi$& $0.141$&$0.072$&$5.31\times 10^{-7}$&$2.64\times 10^{-7}$\\
 $\overline{B}_s\to\rho^0\phi$&
 $0.089$&$0.070$&$1.03\times10^{-6}$&$7.14 \times 10^{-7}$\\\hline
 pure penguin processes\\
 $\overline{B}_s\to K^{0\ast}\overline{K}^{0\ast}$&$0.094$&$0.082$&$2.61\times 10^{-6}$&
 $1.94\times 10^{-6}$\\
 $\overline{B}_s\to K^{0\ast}\phi$& $0.190$&$0.092$&$1.35\times 10^{-7}$&$1.41\times 10^{-7}$\\
 $\overline{B}_s\to \phi \phi$& $0.134$&$0.117$&$1.31\times 10^{-5}$&$9.05\times 10^{-6}$\\
\hline\hline
\end{tabular}
\end{center}
\end{table}
From Table 2, we can find that some channels have large branching
ratios of order $10^{-5}$, which are measurable at near future
experiments at CERN LHCb. Owing to the absence of $(S-P)(S+P)$
penguin operator contributions to $W$-emission amplitudes,
tree-dominated $\bar{B}_s\to V_1 V_2$ decays tend to have larger
branching ratios than the penguin-dominated ones. Moreover, we
find that the transverse to total decay rate $\Gamma_T/\Gamma$ is
very small for most decay modes, so in the heavy quark limit, both
light vector mesons in charmless $\overline{B}_s\to VV$ decays
tend to have zero helicity.

\section{Summary}

In this paper, we calculated the branching ratios for two-body
charmless hadronic $\overline{B}_s\to VV$ decays within the
framework of QCD factorization. Contrary to phenomenological
generalized factorization[9] and NF approach, the nonfactorizable
corrections to each helicity amplitude are not the same. The
effective parameters $a_i^h$ vary for different helicity amplitude
and hence are helicity dependent. Since the leading-twist DAs
contributions to the transversely polarized amplitudes vanish in
the chiral limit, in order to have renormalization scale and
scheme independent predictions, it is necessary to take into
account the contributions of the twist-3 DAs of the vector meson.
Contrary to the $PP$ and $PV$ modes, the annihilation amplitudes
in the $VV$ case do not gain the chiral enhancement of order
$m_B^2/(m_q m_b)$. So we do not include the contributions of the
annihilation diagrams which is truly  power suppressed in the
heavy quark limit. It should be stressed that we have not taken
into account the higher-twist DAs  contribution for the
longitudinally polarized vector meson. Though direct calculation,
the transverse to total decay rate $\Gamma_T/\Gamma$ is found to
be very small, and both light vector mesons tend to have zero
helicity. Branching ratios of $\bar{B}_s\to VV$ decays are
calculated with the LCSR analysis for the form factors and the
branching ratios of some channels are found as large as $10^{-5}$,
which might be accessible at future experiments at CERN LHCb.

\noindent{\bf {\Large Acknowledgments }}

 Y.D is supported by the Henan Provincial Science
Foundation for Prominent Young Scientists under the contract
0312001700. This work is supported in part by National Science
Foundation of China under the contracts 19805015 and 1001750.

\newpage
\noindent{\bf Appendix.}

The $\overline{B_s}\rightarrow VV$ decay amplitudes are collected
here:~~\\
   1. $b\to d$ processes:
 \begin{eqnarray}
 H^h (\overline{B}_s\to K^{+\ast}\rho^-)&=&\frac{G_F}{\sqrt
2}\left\{\lambda_u a_1^h-\lambda_t (a_4^h+a_{10}^h)\right
\}X^{(\overline{B}_s K^{+\ast},\rho^-)}\nonumber\\
 H^h (\overline{B}_s\to K^{0\ast}\rho^0)&=&\frac{G_F}{\sqrt
2}\left\{\lambda_u a_2^h-\lambda_t
(-a_4^h+\frac{3}{2}a_7^h+\frac{3}{2}a_9^h+\frac{1}{2}a_{10}^h)\right
\}X^{(\overline{B}_s K^{0\ast},\rho^0)}\nonumber\\
 H^h (\overline{B}_s\to K^{0\ast}\omega)&=&\frac{G_F}{\sqrt
2}\left\{\lambda_u
a_2^h-\lambda_t (2a_3^h+a_4^h+2a_5^h\right.\nonumber\\
&&\left.+\frac{1}{2}a_7^h+\frac{1}{2}a_9^h-\frac{1}{2}a_{10}^h)\right
\}X^{(\overline{B}_s K^{0\ast},\omega)}.
 \end{eqnarray}
where $\lambda_u=V_{ub} V_{ud}^{\ast}$ and $\lambda_t=V_{tb}
V_{td}^{\ast}$~~\\
   2. $b\to s$ processes:
 \begin{eqnarray}
 H^h (\overline{B}_s\to K^{+\ast}K^{-\ast})&=&\frac{G_F}{\sqrt
2}\left\{\lambda_u a_1^h-\lambda_t (a_4^h+a_{10}^h)\right
\}X^{(\overline{B}_s K^{+\ast},K^{-\ast})}\nonumber\\
 H^h (\overline{B}_s\to \rho^0\phi)&=&\frac{G_F}{\sqrt
2}\left\{\lambda_u a_2^h-\lambda_t
[\frac{3}{2}(a_7^h+a_9^h)]\right
\}X^{(\overline{B}_s \phi,\rho^0)}\nonumber\\
 H^h (\overline{B}_s\to \omega\phi)&=&\frac{G_F}{\sqrt
2}\left\{\lambda_u a_2^h-\lambda_t
[2(a_3^h+a_5^h)+\frac{1}{2}(a_7^h+a_9^h)]\right
\}X^{(\overline{B}_s \phi,\omega)}.
 \end{eqnarray}
where $\lambda_u=V_{ub} V_{us}^{\ast}$ and $\lambda_t=V_{tb}
V_{ts}^{\ast}$~~\\
   3. pure penguin  processes:
\begin{eqnarray}
 H^h (\overline{B}_s\to
K^{0\ast}\overline{K}^{0\ast})&=&-\frac{G_F}{\sqrt 2}V_{tb}
V_{ts}^{\ast} (a_4^h-\frac{1}{2}a_{10}^h)X^{(\overline{B}_s K^{0\ast},\overline{K}^{0\ast})}\nonumber\\
 H^h (\overline{B}_s\to K^{0\ast}\phi)&=&-\frac{G_F}{\sqrt
2}V_{tb}V_{td}^{\ast}\left\{[a_3^h+a_5^h-\frac{1}{2}
(a_7^h+a_9^h)]X^{(\overline{B}_s K^{0\ast},\phi)}\right.\nonumber\\
&&\left.+(a_4^h-\frac{1}{2}a_{10}^h)X^{(B_s \phi,K^{0\ast})}
\right \}\nonumber\\
 H^h (\overline{B}_s\to \phi \phi)&=&-\frac{G_F}{\sqrt
2}V_{tb}V_{ts}^{\ast}[a_3^h+a_4^h+a_5^h-\frac{1}{2}
(a_7^h+a_9^h+a_{10}^h)]X^{(\overline{B}_s \phi,\phi)}.
\end{eqnarray}
In the above expressions, the factorozable amplitude
$X^{(\overline{B}_s V_1,V_2)}$ is defined as in  Eq(\ref{xbs}).
\end{document}